\documentclass[aps,prl,preprint,showpacs,superscriptaddress,groupedaddress]{revtex4}%

\usepackage{graphicx}
\usepackage{dcolumn}
\usepackage{bm}
\usepackage[caption = false]{subfig}
\usepackage{lipsum}
\begin{document}
\title{Angular Dependent Magnetization Dynamics of Kagome Artificial Spin Ice Incorporating Topological Defects }
\author{V.S.~Bhat} \email[]{vinayak.bhat@epfl.ch} \affiliation{\'Ecole Polytechnique F\'ed\'erale de Lausanne, School of Engineering,  Institute of Materials, Laboratory of Nanoscale Magnetic Materials and Magnonics,  1015 Lausanne, Switzerland}\affiliation{Lehrstuhl f\"{u}r Physik funktionaler Schichtsysteme, Physik Department E10, Technische Universit\"{a}t M\"{u}nchen, 85748 Garching, Germany}

\author{F.~Heimbach}\affiliation{Lehrstuhl f\"{u}r Physik funktionaler Schichtsysteme, Physik Department E10, Technische Universit\"{a}t M\"{u}nchen, 85748 Garching, Germany}
\author{I.~Stasinopoulos} \affiliation{Lehrstuhl f\"{u}r Physik funktionaler Schichtsysteme, Physik Department E10, Technische Universit\"{a}t M\"{u}nchen, 85748 Garching, Germany}
\author{D.~Grundler} \email[]{dirk.grundler@epfl.ch}  \affiliation{\'Ecole Polytechnique F\'ed\'erale de Lausanne, School of Engineering,  Institute of Materials, Laboratory of Nanoscale Magnetic Materials and Magnonics,  1015 Lausanne, Switzerland} \affiliation{\'Ecole Polytechnique F\'ed\'erale de Lausanne, School of Engineering,  Institute of Microengineering, Laboratory of Nanoscale Magnetic Materials and Magnonics,  1015 Lausanne, Switzerland}

\vskip 0.25cm
\date{\today}

\begin{abstract}
We report angular-dependent spin-wave spectroscopy on kagome artificial spin ice made of large arrays of interconnected Ni$_{80}$Fe$_{20}$  nanobars. Spectra taken in saturated and disordered states  exhibit a series of resonances with characteristic  in-plane angular dependencies. Micromagnetic simulations allow us to interpret  characteristic resonances of a two-step magnetization reversal of the nanomagnets.  The dynamic properties are consistent with  topological defects that are provoked via a magnetic field applied at specific angles. Simulations that we performed on previously investigated kagome artificial spin ice consisting of isolated nanobars show characteristic discrepancies in the spin wave modes which we explain by the absence of vertices.

\end{abstract}

\maketitle
\section{Introduction}
Spin ice is a geometrically frustrated system where all the competing interactions among spins can not be satisfied at the same time \cite{mengotti2011real}.   Topological defects (TDs) in crystal-based spin ice occur in the form of quasiparticles called monopole-antimonopole (MA) pairs that are separated by a defect string  (Dirac string) \cite{jaubert2009signature,gliga2013spectral}.  Back and forth movements of magnetic monopoles (and hence Dirac strings)  give rise to alternating  \textsc{\char13}magnetic\textsc{\char13}  currents and are promising for magnetronic devices \cite{bramwell2009measurement}.  Techniques that have been used to study these movements can not determine the state of individual spins but rather average over many unit cells. To gain improved statistical insight about frustration, the state of an individual spin should be known. This challenge was accomplished by the material-by-design approach \cite{nisoli2013colloquium} where tailored lattices of Ising spins were constructed in the form of specifically arranged nanobars and studied under various external conditions, such as temperature and magnetic field. In such model systems the occurrence of Dirac monopoles and Dirac strings was studied \cite{mengotti2011real}. One variant  is kagome ASI\textemdash an array of nanobars arranged on a kagome lattice \cite{qi2008direct}. In DC magnetization and simulation studies a two-step magnetization reversal process was observed \cite{burn2015angular,mellado2010kagome,daunheimer2011reducing}. The applied science community is  interested in utilization of ASIs in magnonics related applications \cite{krawczyk2014review}. Indeed, magnonic properties were reported for different ASIs \cite{bhat2016magnetization,zhou2016large,jungfleisch2016dynamic}; however, angular dependencies and the regime of the two-step magnetization reversal were not studied in detail. It is interesting  whether such a two step magnetization process gives rise to distinct resonances, and whether the TDs and their resonances are stable against field cycles. Understanding of their stability and reproducibility is of considerable interest from a fundamental and applied science perspective \cite{gilbert2015ground,libal2012hysteresis,gilbert2015direct}. Here we present spin wave spectra taken for kagome ASI in the saturated state and the hysteretic regime for fields applied under different in-plane angles. We report characteristic resonances that are consistent with the two-step reversal process and are stable under field cycles. We also investigate spin wave spectra for the connected and disconnected kagome ASIs via micromagnetic simulations and encounter characteristic differences. The results are important towards utilization of TDs in kagome lattice as reprogrammable magnonic crystal \cite{krawczyk2014review}.
\section{Experimental Details}
Large arrays (2.4 mm x 2.4 mm) of kagome ASI  [Fig. 1(a)] were fabricated using electron beam lithography and lift-off processing \cite{bhat2016magnetization}. The length $l$, width $w$, and thickness $t$ of a  given Ni$_{80}$Fe$_{20}$ (Py) segment  were kept at 810 nm, 130 nm, and 25 nm, respectively. Room-temperature broadband spin wave spectroscopy measurements  [Fig. 1(b)]  were performed in a flip-chip configuration where a sample was placed face-down on top of a 9 mm long coplanar waveguide. Subsequently, we applied a constant magnetic field (from -1 kOe  to 1  kOe in steps of 0.01 kOe) at a given in-plane angle $\phi$  [Fig. 1(c)]  and performed frequency sweeps from 10 MHz up to 26.5 GHz using a vector network analyzer (VNA). The  spectra presented here correspond to scattering parameters measured by the VNA. We used the difference technique to increase the signal-to-noise ratio in that we subtracted a reference spectrum \cite{bhat2016magnetization}.\\ Simulations were performed on a kagome ASI system consisting of 132 nanobars  [Fig. 1(d)] using the OOMMF code \cite{OOMMF1}. The parameters used in simulations were as follows: Exchange constant $A = 1.3 \times 10^{-11} \,\mathrm{J\,m^{-1}}$, saturation magnetization $M_{\mathrm{S}}= 7\times 10^{5} \,\mathrm{A\,m^{-1}}$ \cite{wang2007geometrical}, magnetocrystalline anisotropy constant $K = 0$, gyromagnetic ratio $\gamma = 2.211 \times 10^{5} \,\mathrm{m\,A^{-1}\,s^{-1}}$, and damping coefficient $\alpha = 0.01$.  We extracted spatial distributions ($x,y$ maps) of magnetization vectors, demagnetization fields, and spin-precessional amplitudes. Different magnetic configurations were explored by rotating the sample by a given degree, initializing magnetization vectors of individual segments \cite{gliga2013spectral} and then relaxing the spin system at the given magnetic field $\bf{H}$ applied along the $x-$axis. For this, we first created a colored bitmap using a grid of 10 nm$\times$10 nm$\times$25 nm where each segment was assigned a color corresponding to its assumed magnetization orientation in the reference state.  Our reference state corresponded to a field-driven remanent state, i.e., the sample was initially saturated along the $x-$direction, and the field was then reduced to zero adiabatically. This way all of the segments (nanobars) in the reference state had their magnetization direction along their long axis, and each segment displayed a positive component of magnetization along the $+x-$direction. Following Refs. \cite{gliga2013spectral,bhat2016magnetization} we then introduced MA pairs separated by a Dirac string by assigning a different color  [Fig. 1(d)] to relevant segments (which corresponded to the magnetization rotated by 180$^{\circ}$). Subsequently, we imported this colored bitmap into OOMMF and equilibrated the magnetic configuration in the presence of a field that resided within the experimental switching regime. A dynamical simulation was conducted by applying a Gaussian magnetic field pulse of full width at half maximum of 2.5 ps and a strength of 0.2 kOe perpendicular to the film plane. We logged 256 values of the time-dependent magnetization vector for each grid pixel at 20-ps time steps. The perpendicular component of magnetization was recorded as a function of $x, y$ and the time step. A fast Fourier transformation (FFT) was performed on the magnetization of each pixel along the time axis to obtain the resonance spectrum. The absorbed power for each pixel was obtained by squaring the FFT amplitude, and the signal phase was obtained by calculating the imaginary part of the amplitude. The analysis yielded an areal map of the spin-precessional amplitudes at the specific frequency. The power spectra were obtained by integrating over all pixels for each frequency step. We applied the charge model for classifying  different magnetic and charge configurations in the studied kagome ASI \cite{castelnovo2008magnetic,mellado2010dynamics,mengotti2011real}. We considered each Py segment to be a dumbbell with charges $+q\; (= M \times t \times w \times l/l)$ and $-q $ at its opposite ends  [Fig. 1(a)], where $ M $ represented the magnetization of a Py nanomagnet. Each vertex (excluding the boundary) in a kagome ASI possessed a coordination number of $ 3 $. The well-defined and reproducible reference state was represented by a lattice where each vertex in the kagome ASI carried a total charge of $Q = +q$ or $Q = -q$. A monopole (antimonopole) occured when $Q_{f} - Q_{i} = \Delta Q > 0 $ ($Q_{f} - Q_{i} = \Delta Q < 0 $), where $ Q_{i}$ and $Q_{f} $ corresponded to the total charge of a given vertex before and after the reversal of a segment connected to it, respectively \cite{mellado2010dynamics,mengotti2011real}.  
\section{Results and Discussions}
Figure 1(b) shows spectra taken at $ H = 1$ kOe for different field orientations $\phi$ between 0$^{\circ}$ and $\phi$ = 32$^{\circ}$. At $\phi$ = 0$^{\circ}$ (bottom curve), we find two prominent modes $C$ and $D$  (minima at $f$ = 9.7 and 12.6 GHz)  and two weak ones (modes $A$ at $f$ = 5.6 and $B$ at $f$ = 7.5 GHz)  consistent with Ref. \cite{bhat2016magnetization}. As $\phi$ is increased, the mode $C$ splits into two modes $C1$ and $C3$; at $\phi$ = 30$^{\circ}$ modes $C3$ and $D$ merge into each other and mode $C1$ resides at the lowest frequency. This behavior is best seen in  Fig. 1(c) in which we summarize spectra for $H=1$ kOe and different $\phi$ in a gray-scale plot (bottom) and extract characteristic branches (top) as indicated by the different colors. The bottom panel displays the variation of spectra with $\phi$ in that we subtracted the spectra from two subsequent angles $\phi$ and plot this difference as a function of $\phi$ (first derivative with respect to $\phi$). Apart from angles $\phi=0, \pm60, \pm 120$ and $\pm 180^{\circ}$ we observe three prominent branches with different eigenfrequencies. Assuming that $\mathbf{M}$ of each nanobar is nearly aligned with $\mathbf{H}$ at the relatively large field of 1 kOe the three different eigenfrequencies reflect the three different orientations T1 to T3 [Fig. 1 (d)] that segments of the kagome lattice experience with $\mathbf{H}$. Due to the correspondingly different demagnetization field that enters the total effective field $H_{\rm eff}$ and the equation of spin-precessional motion \cite{gurevich1996magnetization} three different eigenfrequencies are expected for segments T1 to T3 whenever $\mathbf{H}$ does not point along a high symmetry direction. A maximum eigenfrequency is detected whenever the field $\mathbf{H}$ is collinear with both a high symmetry direction and a specific set T$_i$ ($i=1,2,3$) of segments of identical spatial orientation. In Ref. \cite{bhat2016magnetization} it was shown that TDs in a kagome ASI locally modify $H_{\rm eff}$ leading to characteristic modifications in eigenfrequencies $f$ for individual segments. We will exploit this feature when analyzing spin-wave spectra in the hysteretic regime. We note that compared to previous studies on disconnected kagome ASI \cite{zhou2016large}, the slopes $ |df/d \phi| $ of the three branches in Fig. 1(c) are higher, substantiating a higher shape anisotropy in the kagome ASI that we studied; in our case, segments are only 130 nm wide as opposed to 230 nm in the case of Ref. \cite{zhou2016large}. \\
Before we discuss the detailed field and angular dependencies of the different resonances, we explore their reproducibility in the hysteretic regime. We took data in minor loops at $\phi = 0^{\circ}$ where we expected the generation and existence of TDs \cite{bhat2016magnetization}.  First, we saturated the sample along $\phi = 0^{\circ}$ at positive $H$ and then entered the hysteretic regime by applying a negative $H=-H_{\rm rev}$ where $H_{\rm rev}$ was smaller than the saturation field of 0.46 kOe [Fig. 2]. We then collected spin wave for fields between $H_{\rm rev}$ and $-H_{\rm rev}$ (minor loop).  In Fig. 2 (a),  spin wave spectra collected for a field sequence of $H=-H_{\rm rev}=-0.2 $ kOe  $ \rightarrow $  $ 0   $  kOe  $ \rightarrow $   $ +0.2  $ kOe  $ \rightarrow $  $ 0 $  kOe  $ \rightarrow $  $ -0.2 $  kOe  $ \rightarrow $  $ 0 $  kOe  $ \rightarrow $   $ +0.2 $  kOe are shown. In Fig. 2(b) and (c) larger values for $H_{\rm rev}$ were used. Spectra were found to be reproducible when we returned to  the same field. This behavior suggested that TDs in kagome ASI generate stable spectra in a minor loop cycle.  \\
We now discuss spectra taken at fixed angles $\phi \neq 0^{\circ}$ over a broad range of $H$ including both the saturated and hysteretic regime. In Fig. 3, we summarize the eigenfrequencies (symbols) of spin-wave modes that we extracted from spectra obtained for $\phi = 15^{\circ}$ [Fig. 3(a)], $ 20^{\circ}  $ [Fig. 3(b)],  and $ 30^{\circ} $ [Fig. 3(c)]. We always varied the applied field from +1 kOe to -1 kOe in a step-wise manner. The lines represent simulated field dependencies of eigenfrequencies for saturated ASIs (no TDs). The dynamic response of the ASI is found to change considerably for different $\phi$. In the following, we list the most prominent features.  At $\phi$ = 15$^{\circ}$ [Fig. 3(a)],  five  branches ($A$, $C1$, $C2$, $C3$ and $D$) are found in the  saturated state at large positive $H$. As the applied field is reduced, the five modes approached each other. At $H$ = 0 kOe, only branch $D$ was clearly resolved.  At $ H $ = -0.3 kOe, a branch reappeared that exhibited a steep slope $df/dH<0$ consistent with the reversal of nanomagnets introducing TDs. The amplitude of this high frequency mode grew with decreasing field indicating the switching of more and more segments. At the same time the mode at the lower frequency successively vanished. At $H= -0.44$ kOe and $-0.47$ kOe two further modes appeared, which grew stronger at more negative $H$.  We introduced vertical broken arrows and labels 1 to 3 when irreversible changes appeared in the field dependencies of branches. Later, we will show that these changes reflect the reversal of $T1$  to $T3$ segments, respectively [Fig. 1(d)]. At  $\phi$ = 20$^{\circ}$ [Fig. 3(b)] again  five  branches ($A$, $C1$, $C2$, $C3$ and $D$) were found at $H$ = 1 kOe; here, the frequency separation between $C3$ and $D$ was smaller compared to Fig. 3(a). The overall field dependencies of the branches were similar to Fig. 3(a); however, all the reversals of segments occurred at more negative fields [compare broken vertical arrows in Figs. 3(a) and 3(b)]. The resonance field values for branches $C1$, $C3$ and $D$ at $\phi$ = 20$^{\circ}$ during the magnetization reversal were larger by 0.12, 0.16, and 0.01 kOe, respectively, as compared to that  for $\phi$ = 15$^{\circ}$, which indicated a dependence of TD formation on the applied field angle. At  $\phi$ = 30$^{\circ}$ [Fig. 3(c)], two out of only three branches were prominent. The prominent branches are labelled by $ C1 $ and $ D $. The former branch $ C3 $ became degenerate with $ D $, and they merged to a single branch labelled by $D$. While branch $D$  displayed a smaller slope $|df/dH|$ compared to smaller $\phi$, the slope of branch $C1$ was much larger than before and at its maximum value in Fig. 3.
\subsection{Simulations and Analysis}
To understand in detail the origin of the different branches in Fig. 3, we performed simulations on interconnected kagome ASI in which we considered different angles $\phi$ (spectra represented by solid lines in Fig. 4). For $\phi$ =  15$^{\circ}$, we categorize the segments based on their angles with respect to the applied field [Fig. 1(d)].       Simulations  at $ H = 1$ kOe provided three prominent modes at $f$ = 8.6 (mode $C1$), 11 (mode $C3$), and 12.6 GHz (mode $D$). Maps of spin precessional motion indicated that modes $ C1 $, $ C3 $, and $ D $  originated from $ T1 $, $ T2 $, and $ T3 $ segments, respectively.  At $ H = -0.3$ kOe  two broad modes can be seen in Fig. 4a;  these  modes represent power absorption in all three types of segments. Here, TDs are not yet present. Previous DC magnetization studies showed that  ferromagnetic reversal in kagome ASI took place in two steps for $  \phi = 20 ^{\circ}$ \cite{mellado2010dynamics,daunheimer2011reducing}. Switching was found to begin in $T2$ and $T3$ segments, followed by the reversal of $T1$ segments at more negative $H$. Following these earlier findings, we introduced different numbers of TDs in the form of MA pairs and Dirac strings by first reversing $T2$ and $T3$ segments, whose eigenfrequencies were degenerate. For 7  MA pairs [Fig. 4(a)], we found the appearance of an additional high-frequency mode that we labelled with $D$. Local power maps indicated the power absorption in reversed $T2$ and $T3$ segments with the degenerate eigenfrequencies. Further introduction of TDs caused the amplitude of this mode to increase at the expense of a mode that was due to non-reversed segments (compare spectrum with 17 MA pairs). As  $T1$ segments reversed, we observed a mode $C3$ developing close to mode $D$. For $\phi$ = 20$^{\circ}$ micromagnetic simulations  at $H$ = 1 kOe showed three prominent modes [Fig. 4(b)] at $f$ = 8.1 (mode $C1$), 11.2 (mode $C3$), and 12.4 GHz (mode $D$)  that originated from $T1$, $T2$, and $T3$ segments, respectively. Compared to $\phi$ = 15$^{\circ}$, we observed that modes $C1$, $C3$, and $D$  were at lower, higher and lower frequencies, respectively, consistent with experimental results [Fig. 1(b)]. At $H$ = -0.3 kOe when TDs were not present, two broad modes at $f$ = 7.52 and 9.7 GHz were found for $\phi$ = 20$^{\circ}$ [Fig. 4(b)]; local power maps indicated that these were degenerate modes representing power absorption in all thee types of segments. When we introduced TDs by reversing $T2$ and $T3$ segments for this angle we observed the appearance of mode $D$ at $f$ = 9.9 GHz in the spectrum labelled as MA 4 (i.e., 4 MA pairs). Once we reversed $T1$ segments, we observed a mode evolving out of the broad high frequency mode. Thus simulations on $\phi$ = 15$^{\circ}$ and 20$^{\circ}$  showed a systematic trend in the occurrence of this peak $C3$. In our experiments, we observed a branch to develop at more negative $H$ compared to mode $C1$. This was in contrast to previously published data taken at $\phi$ = 0$^{\circ}$ only \cite{bhat2016magnetization}.
\\ \indent To obtain detailed insight behind emergence of these spin wave branches, we studied the local spin-precessional amplitudes (power absorption) at a given resonance frequency.  Simulations show that the local power absorption  strongly depends on the local demagnetizing field and in turn on the total effective field [Fig. 5(a) and (b)].  Local power maps showed that branch $D$ was due to reversed $T3$ segments and reversed $T2$ segments that {\em were not} surrounded by MA pairs [Fig. 5(c)]. The spin wave mode corresponding to branch $C3$ was due to power absorption in reversed segments that {\em were} surrounded by MA pairs [Fig. 5(d)]. The branch $C1$  appeared when $T1$ segments were reversed. We found that the resonance frequency of $T2$ segments surrounded by MA pairs was lower as compared to the ones that were on a Dirac string. We explain this feature by the inhomogeneous demagnetizing field. The local demagnetizing field in $T2$ segments that are surrounded by reversed $T1$ and $T3$ segments points against  the applied field direction; therefore, the total effective field $H_{\rm eff}$ is small and spins resonate at lower frequency in the $T2$ segments surrounded by non-reversed $T1$ segments. \\
The sudden emergence of mode $C3$ in the experiment and simulation after switching of $T1$ segments points towards the presence of two-step magnetization reversal process in the studied kagome ASI that we will discuss in the following based on a comparison of Figs. 3(a), 3(b), 4, and 5. The magnetization reversal in kagome ASI begins at $H$ = -0.3 kOe (-0.31 kOe) for $\phi$ = 15$^{\circ}$ ($\phi$ = 20$^{\circ}$) by switching of $T2$ and $T3$ segments, which is indicated by an arrow labelled $1$ in Figs. 3(a) and 3(b). The changes in slope $df/dH$ of branch $D$ as marked by arrows with label $2$ in Figs. 3(a) and 3(b) indicate that the switching of $T2$ and  $T3$ segments ends at $H$ = -0.35 kOe (-0.38 kOe) for  $\phi$ = 15$^{\circ}$ ($\phi$ = 20$^{\circ}$). We then observe further a change in slope in branch $D$ coinciding with the appearance of branches $C1$ and $C3$ at $H$ = -0.44 kOe (-0.58 kOe) for $\phi$ = 15$^{\circ}$ ($\phi$ = 20$^{\circ}$) [indicated by arrows with label $3$ in Figs. 3(a) and 3(b)].  For an ASI with interconnected nanobars we have thus found that the two-step reversal process is accompanied by  both the appearance of novel branches and changes in $df/dH$ of branches already present.\\
Previous quasi-static as well as magnetodynamic studies have been focussed mainly on either a connected or disconnected network of kagome ASIs \cite{qi2008direct,mellado2010kagome}; however, a direct comparison between these two types of systems is still lacking. We compared the magnetodynamics signatures of connected and disconnected networks of Py nanobars arranged on a kagome lattice via micromagnetic simulations. For the disconnected ASI, we fixed the length, width, and thickness of a given Py nanobar to 410 nm, 130 nm, and 25 nm, respectively with the same lattice constant as in the case of interconnected kagome ASI [see dotted lines in Fig. 4]. Simulations in the saturated and hysteretic regimes showed that spin wave resonances were lower in amplitude and occurred at lower frequency values as compared to connected kagome ASI. As we introduced TDs by switching selected $T1$ nanobars, the local power maps for disconnected kagome ASI  showed similar resonance frequencies for $T2$ nanobars that were surrounded by either switched or unswitched $T1$ nanobars. This result is in contrast to the ASI with interconnected nanobars for which we found different eigenfrequencies. This finding indicates that the vertices of interconnected nanobars play a major role for the local modification of resonance frequencies $f$ due to TDs. 

\section{Summary}
To summarize, we measured spin wave spectra of interconnected kagome ASI prepared from Py nanobars in an external field applied at different in-plane angles in the hysteretic and saturated regimes. We observed magnetodynamics signatures of the two-step magnetization reversal process.  Experimental spectra and simulations identify a specific resonance that emerges on the low-frequency side of a broad spin-wave mode to indicate the onset of the 2nd step of the magnetization reversal process of kagome ASI. By comparison between interconnected and disconnected kagome ASIs via micromagnetic simulations we argue that the vertices in interconnected ASI are key to observe significant resonance frequency modifications induced by topological defects. At the same time, interconnected nanobars allow one to transmit and manipulate exchange-dominated spin waves of short wavelengths. Adjustable resonance frequencies that are modified by the TDs are at the heart of reprogrammable magnonic crystals.

  \begin{figure}
  	\includegraphics[width=0.8\textwidth]{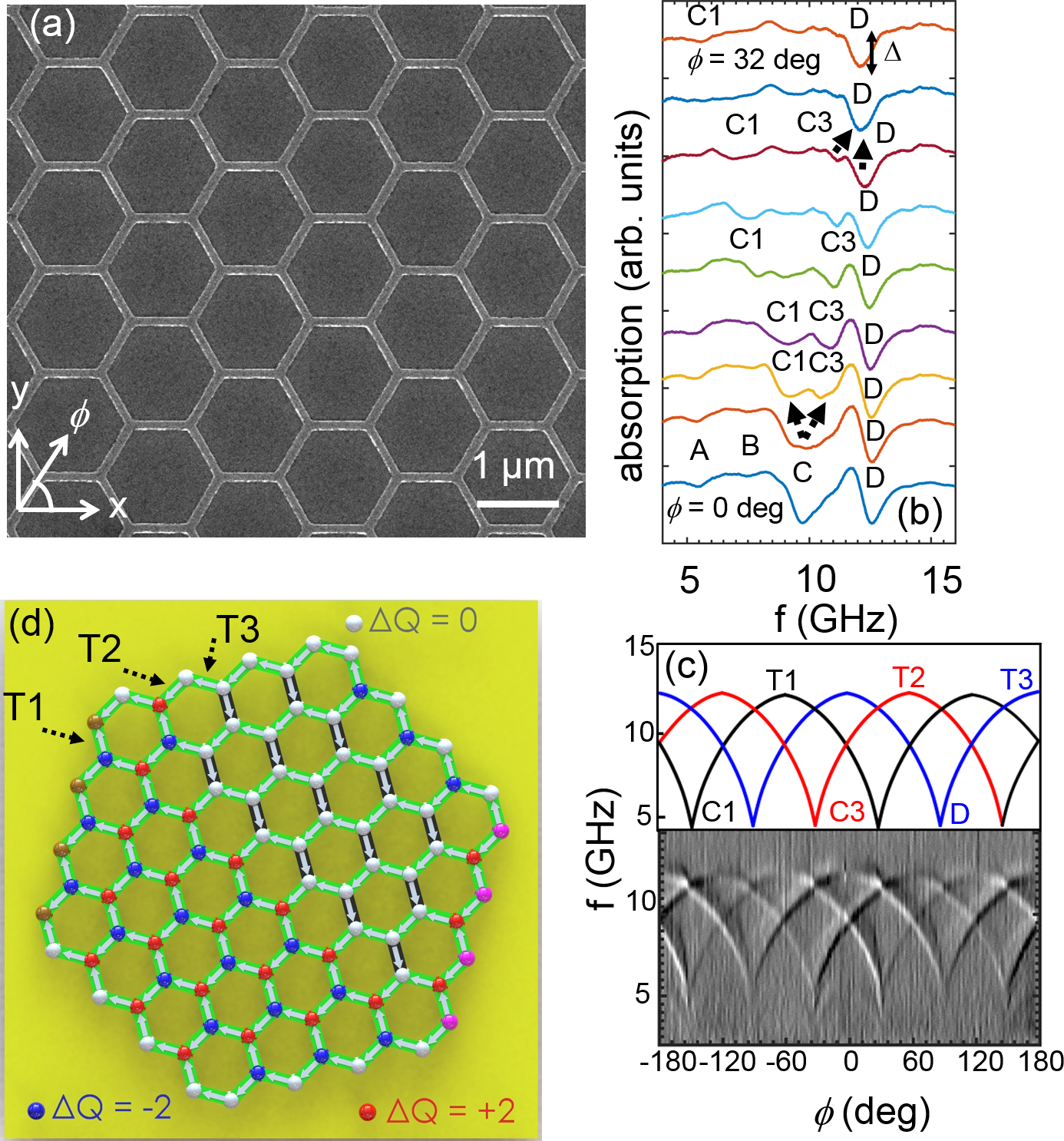}
  	\begin{flushleft}
  		\caption{(a) Representative SEM image of the kagome ASI sample. Bright region corresponds to Py and dark region represents GaAs substrate. (b) Spin wave spectra at different in-plane angles of the applied field. The mode shown underneath the label \textit{C} splits into two modes as $\phi$ increases (indicated by arrows). The definition of amplitude variation ($\Delta$) can also be seen. (c) Gray-scale  plot  depicting spin wave absorption spectra   as a function of in-plane angle $ \phi $  (bottom)  and the cartoon representation of  extracted branches (top). The branches corresponding to modes $C1$ (nanobars of type $T1$), $C3$ (nanobars of type $T2$), and $D$ (nanobars of type $T3$) are represented by black, red, and blue color lines, respectively, in the cartoon representation.  (d)  Charge model representation of the kagome ASI incorporating Dirac strings with MA pairs for $\phi=15^{\circ}$.  Red, dark yellow, blue, magenta and gray spheres represent monopoles, doubly-charged monopoles, antimonopoles, doubly-charged antimonopoles and reference configuration, respectively. The dashed black color lines indicate nanobars of type $T1$, $T2$, and $T3$.  }
  	\end{flushleft}
  \end{figure}
\begin{figure}
	\includegraphics[width=0.9\textwidth]{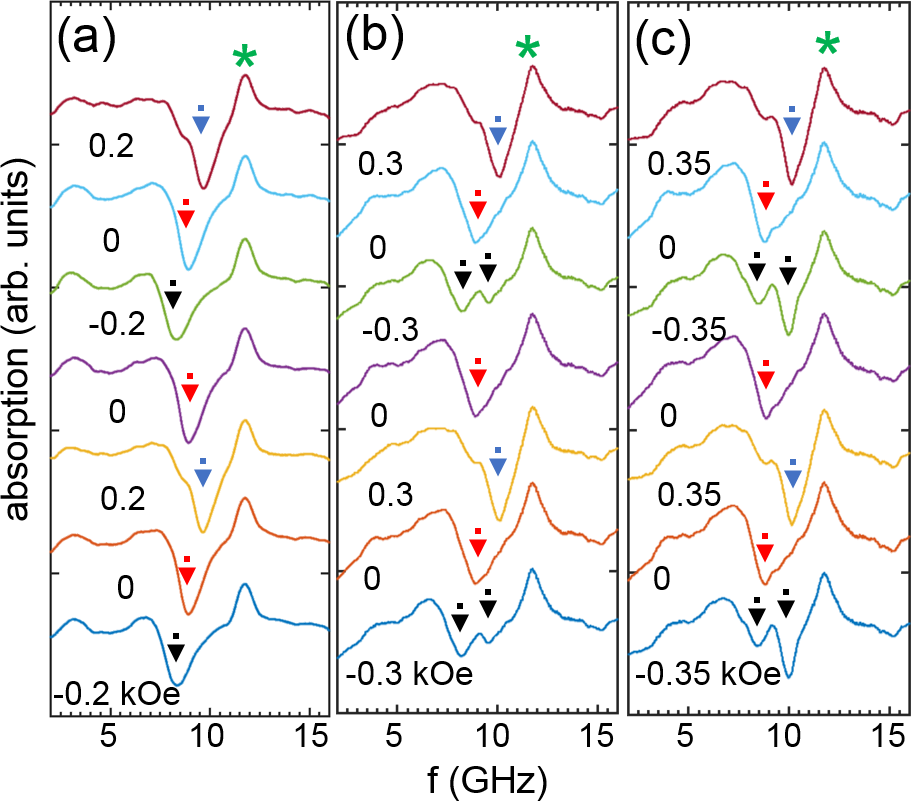}
	\begin{flushleft}
		\caption{ (a), (b), and (c) Spin wave spectra for different external fields in kOe (labels) at $\phi=0^{\circ}$. Numerical values above each curve denote respective field values in kOe. The arrows indicate relevant resonances. The asterisk symbol \textsc{\char13}*\textsc{\char13} marks a feature that represents a resonance in the reference spectrum and appears due to the difference technique.}
	\end{flushleft}
\end{figure}
\begin{figure}	
	\includegraphics[width=0.84\textwidth]{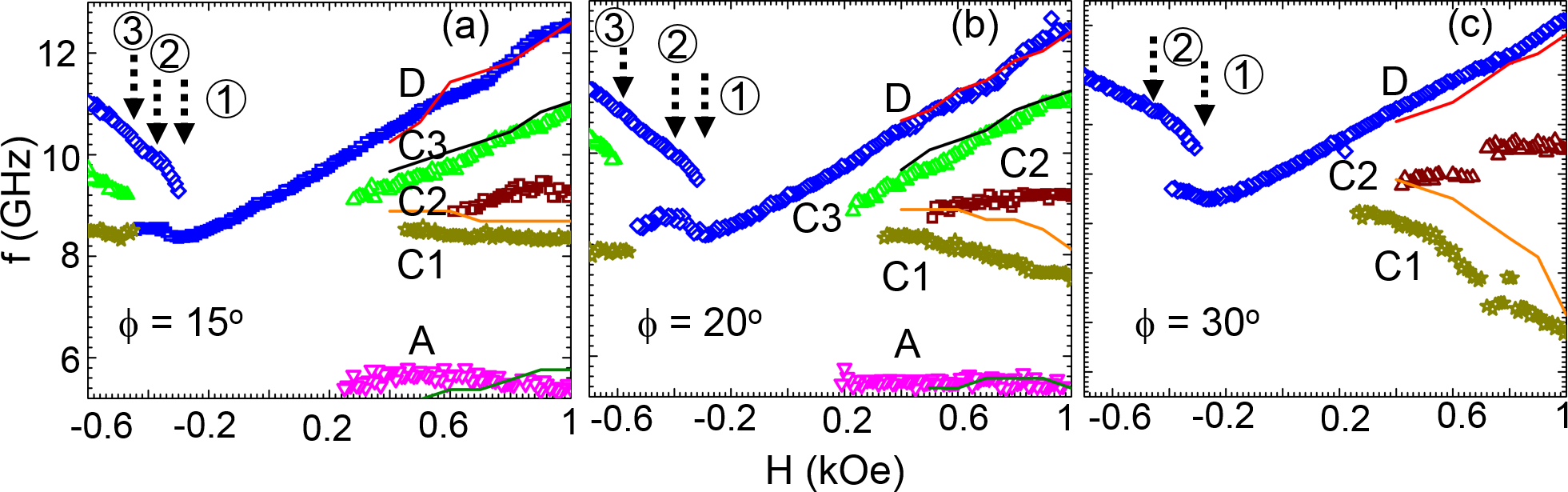}
	\begin{flushleft}
		\caption{ Experimental (symbols) and simulated (lines) resonance frequencies $f$ obtained for $H$ decreasing from 1~kOe for (a) $\phi = 15^{o}$ (b) $\phi = 20^{o}$ and (c) $\phi = 30^{o}$. The symbols size indicates the error bar in $f$. The black color dashed vertical arrows labeled $ 1 $ and $2$ mark the onset and end, respectively, of switching of $T3$ and $T2$ segments. The arrow $ 3 $ indicates the onset of switching in $T1$ nanobars.  }
	\end{flushleft}
\end{figure}
   \begin{figure}	
   	\includegraphics[width=0.8\textwidth]{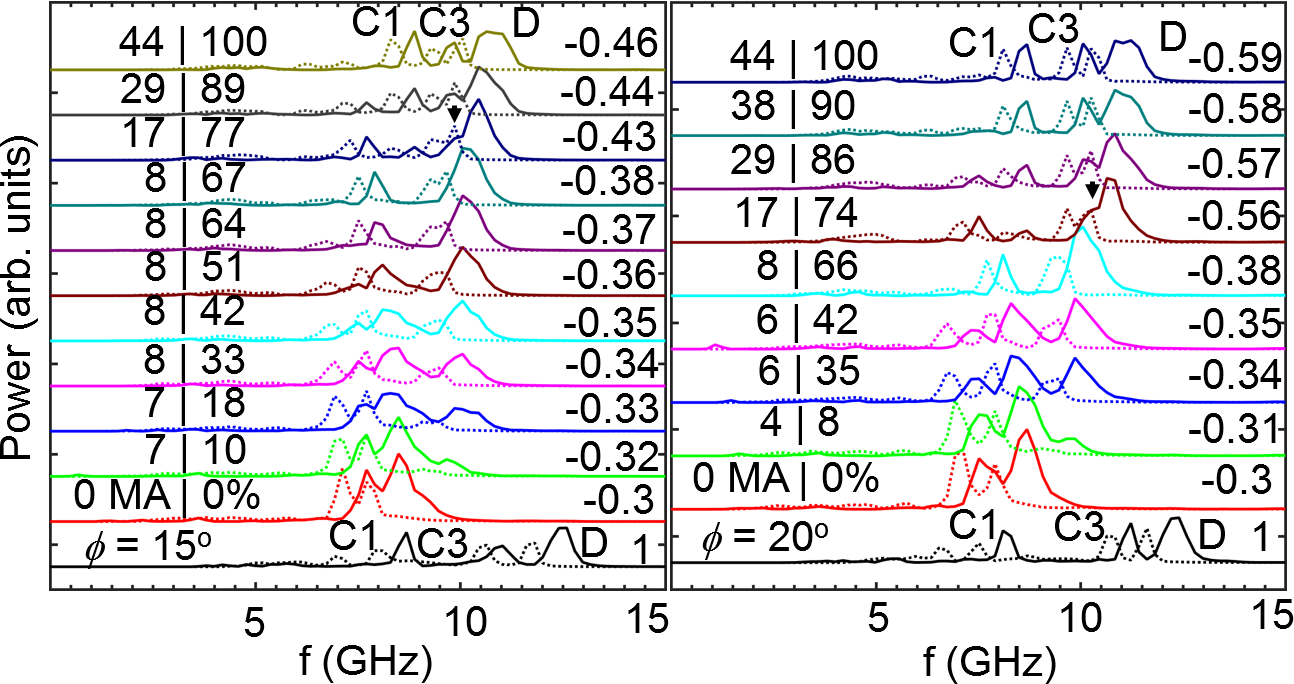}
   	\begin{flushleft}
   		\caption{ Simulated power spectra for the saturated state at $H=1$~kOe, -0.3~kOe, and for different TDs configurations at successively decreased $H$ (from bottom to top) (a) $\phi=15^{\circ}$  and (b) $\phi=20^{\circ}$. Solid and dotted lines represent simulation results for connected and disconnected kagome ASIs, respectively.  Numbers on the left and right side of $\textquotedblright |\textquotedblright$ indicate the number $n$ of monopole-antimonopole pairs and percentage of nanobars switched, respectively. The vertical black color arrows mark the additional mode that emerges on the low-frequency side of the broad resonance that indicates the onset of reversal of $T1$ segments in interconnected kagome ASI (spectra shown as solid lines). The corresponding mode is not detected in spectra simulated for disconnected kagome ASI (dotted lines). Numbers on the right indicate field values in kOe. }
   	\end{flushleft}
   \end{figure}
   \begin{figure}
   	\includegraphics[width=0.95\textwidth]{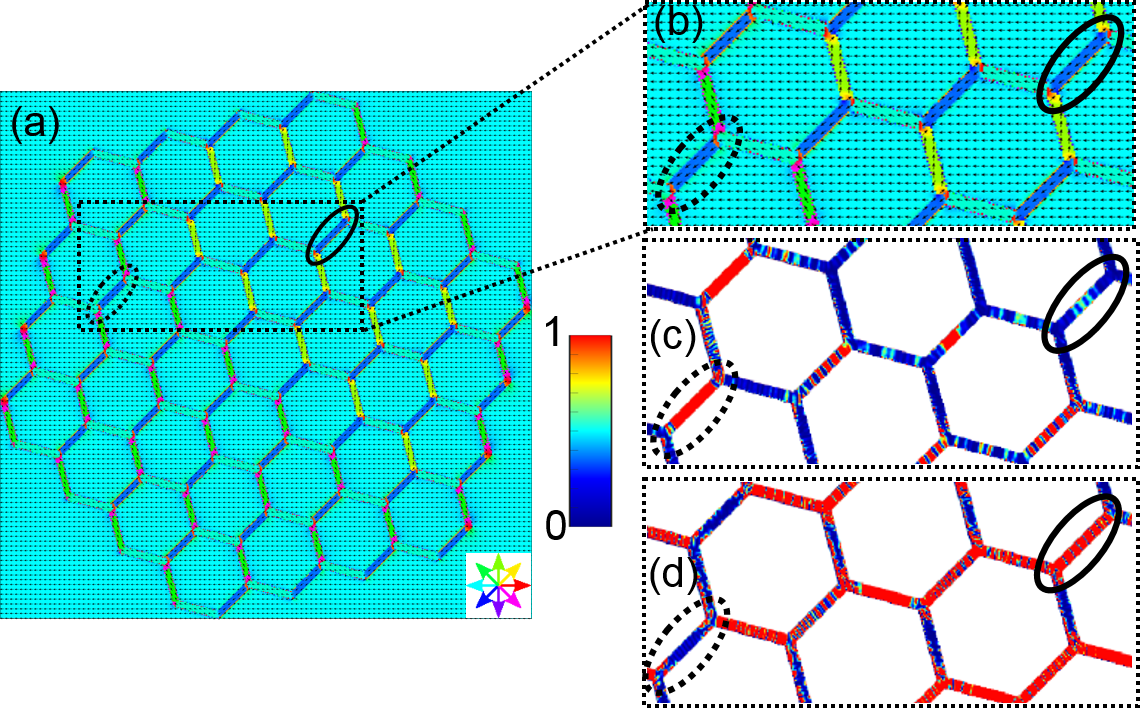}
   	\begin{flushleft}
   		\caption{(a) Directions of the total effective field $ H_{\rm eff} (x,y) $ (color wheel) at $ H $ = -0.44 kOe and  $\phi=15^{\circ}$  for the ASI state shown in Fig. 1(d). (b) Enlarged view of a section indicated by the dotted box in (a). The magnitude of $ H_{\rm eff}(x,y) $  in the center of  $T2$ nanobars highlighted by dotted and solid ellipses is $ 2.38 \times 10^{4} $ A/m and $ 2.74 \times 10^{4} $ A/m, respectively.   Local power maps illustrating spin-precessional motion (color code shown in the legend) at (c) 10.25~GHz (showing main power absorption in  $T2$ segments that are on Dirac strings) and (d) 9.7~GHz (showing main power absorption in $T2$ segments confined by MA pairs).  The different values of $ H_{\rm eff} (x,y) $  found in (b) are consistent with the difference in the spin wave frequency for $T2$ nanobars in (c) and (d).      }
   	\end{flushleft}
   \end{figure}
\indent  The research was supported by the SNSF via grant number 163016 and Transregio TRR80 'From electronic correlations to functionality' via the DFG.

\end{document}